\documentclass[twocolumn]{aastex701}

\usepackage{amsmath,amsfonts,amssymb,mhchem}
\usepackage{xspace}
\usepackage{hyperref}
\usepackage{graphicx}
\usepackage{natbib}
\usepackage[most]{tcolorbox}
\usepackage{wasysym}
\definecolor{darkred}{rgb}{0.5, 0.0, 0.0}

\usepackage{siunitx}

\begin{document}

\title{Leveraging Impact Parameter to Mitigate the Transit Light Source Effect: \\Early Insights from TRAPPIST-1}

\correspondingauthor{Ana Glidden}
\email{aglidden@mit.edu}

\author[0000-0002-5322-2315]{Ana Glidden}
\affiliation{Department of Earth, Atmospheric and Planetary Sciences, Massachusetts Institute of Technology, Cambridge, MA 02139, USA}
\affiliation{Kavli Institute for Astrophysics and Space Research, Massachusetts Institute of Technology, Cambridge, MA 02139, USA}
\email{aglidden@mit.edu}

\author[0000-0002-8842-5403]{Alexander I. Shapiro}
\affiliation{Institute of Physics, University of Graz, A-8010 Graz, Austria }
\affiliation{Max-Planck-Institut f\"ur Sonnensystemforschung, Justus-von-Liebig-Weg 3, 37077 G\"ottingen, Germany}
\email{alexander.shapiro@uni-graz.at}

\author[0000-0002-6892-6948]{Sara Seager}
\affiliation{Department of Physics and Kavli Institute for Astrophysics and Space Research, Massachusetts Institute of Technology, Cambridge, MA 02139, USA}
\affiliation{Department of Earth, Atmospheric and Planetary Sciences, Massachusetts Institute of Technology, Cambridge, MA 02139, USA}
\affiliation{Department of Aeronautics and Astronautics, Massachusetts Institute of Technology, 77 Massachusetts Avenue, Cambridge, MA 02139, USA}
\email{seager@mit.edu}

\author[0000-0002-6087-3271]{Nadiia Kostogryz}
\affiliation{Max-Planck-Institut f\"ur Sonnensystemforschung, Justus-von-Liebig-Weg 3, 37077 G\"ottingen, Germany}
\email{kostogryz@mps.mpg.de}

\author[0009-0009-3020-3435]{Valeriy Vasilyev}
\affiliation{Max-Planck-Institut f\"ur Sonnensystemforschung, Justus-von-Liebig-Weg 3, 37077 G\"ottingen, Germany}
\email{vasilyev@mps.mpg.de}

\author[0000-0001-7827-7825]{Roeland P. van der Marel}
\affiliation{Space Telescope Science Institute, 3700 San Martin Drive, Baltimore, MD 21218, USA}
\affiliation{William H. Miller III Department of Physics and Astronomy, Johns Hopkins University, Baltimore, MD 21218, USA}
\email{marel@stsci.edu}

\author[0000-0003-2415-2191]{J.~de~Wit}
\affiliation{Department of Earth, Atmospheric and Planetary Sciences,
Massachusetts Institute of Technology, Cambridge, MA 02139, USA}
\email{jdewit@mit.edu}

\author[0000-0002-3627-1676]{Benjamin V.\ Rackham}
\affiliation{Department of Earth, Atmospheric and Planetary Sciences, Massachusetts Institute of Technology, 77 Massachusetts Avenue, Cambridge, MA 02139, USA}
\affiliation{Kavli Institute for Astrophysics and Space Research, Massachusetts Institute of Technology, Cambridge, MA 02139, USA}
\email{brackham@mit.edu}

\author[0000-0002-8052-3893]{Prajwal Niraula}
\affiliation{Department of Earth, Atmospheric and Planetary Sciences, Massachusetts Institute of Technology, Cambridge, MA 02139, USA}
\affiliation{Kavli Institute for Astrophysics and Space Research, Massachusetts Institute of Technology, Cambridge, MA 02139, USA}
\email{pniraula@mit.edu}

\author[0000-0002-0832-710X]{Natalie H. Allen}
\affiliation{William H. Miller III Department of Physics and Astronomy, Johns Hopkins University, Baltimore, MD 21218, USA}
\email{nallen19@jhu.edu}

\author[0000-0001-5732-8531]{Jingcheng Huang}
\affiliation{Department of Earth, Atmospheric and Planetary Sciences, Massachusetts Institute of Technology, Cambridge, MA 02139, USA}
\email{huangjc@mit.edu}

\author[0000-0002-8507-1304]{Nikole K. Lewis}
\affiliation{Department of Astronomy and Carl Sagan Institute, Cornell University, 122 Sciences Drive, Ithaca, NY 14853, USA}
\email{nkl35@cornell.edu}

\author[0000-0003-0525-9647]{Zifan Lin}
\affiliation{Department of Physics and McDonnell Center for the Space Sciences, Washington University, St. Louis, MO 63130, USA}
\email{lzifan@wustl.edu}

\author[0000-0002-0746-1980]{Jacob Lustig-Yaeger}
\affiliation{Johns Hopkins Applied Physics Laboratory, Laurel, MD 20723, USA}
\email{Jacob.Lustig-Yaeger@jhuapl.edu}

\author[0000-0003-4816-3469]{Ryan J. MacDonald}
\affiliation{School of Physics and Astronomy, University of St Andrews, North Haugh, St Andrews, KY16 9SS, UK}
\email{Ryan.MacDonald@st-andrews.ac.uk}

\author[0000-0003-2528-3409]{Brett M. Morris}
\affiliation{Space Telescope Science Institute, 3700 San Martin Dr, Baltimore, MD 21218, USA}
\email{bmmorris@stsci.edu}

\author[0000-0003-0814-7923]{Elijah Mullens}
\affiliation{Department of Astronomy and Carl Sagan Institute, Cornell University, 122 Sciences Drive, Ithaca, NY 14853, USA}
\email{eem85@cornell.edu}

\author[0000-0002-7352-7941]{Kevin B. Stevenson}
\affiliation{Johns Hopkins APL, 11100 Johns Hopkins Rd, Laurel, MD 20723, USA}
\email{kevin.stevenson@jhuapl.edu}

\author[0000-0003-3305-6281]{Jeff A. Valenti}
\affiliation{Space Telescope Science Institute, 3700 San Martin Drive, Baltimore, MD 21218, USA}
\email{valenti@stsci.edu}

\author[0000-0002-2643-6836]{Daniel Valentine}
\affiliation{University of Bristol, HH Wills Physics Laboratory, Tyndall Avenue, Bristol, UK}
\email{daniel.valentine@bristol.ac.uk}

\author[0000-0003-4328-3867]{Hannah R. Wakeford}
\affiliation{School of Physics, University of Bristol, HH Wills Physics Laboratory, Tyndall Avenue, Bristol BS8 1TL, UK}
\email{hannah.wakeford@bristol.ac.uk}

\author{C. Matt Mountain}
\affiliation{Association of Universities for Research in Astronomy, 1331 Pennsylvania Avenue NW Suite 1475, Washington, DC 20004, USA}
\email{mmountain@aura-astronomy.org}

\begin{abstract}

Stellar activity complicates exoplanet transmission spectra, particularly for smaller planets around M dwarfs with JWST. The transit light source (TLS) effect, the imprinting of spectral differences between the average stellar disk and the occulted transit chord onto the transmission spectrum, makes it challenging to directly use the out-of-transit spectrum to correct for stellar contamination. Theory and observations suggest that spots may concentrate towards higher latitudes when the Coriolis force is substantial relative to buoyancy, leaving the equatorial region relatively quiet. Here, we evaluate how the latitudinal distribution of active regions shapes the strength of the TLS effect for planets spanning a range of impact parameters ($b$), using TRAPPIST-1 as a testbed. We first construct a fiducial model to illustrate two distribution regimes. With our model, the moderate-$b$ outer TRAPPIST-1 planets (f, g, h) occult a more typical region of the stellar disk than the inner planets and are thereby less affected by the TLS effect, though their bias may vary more from visit to visit as these active regions evolve with time. More generally, our results imply an impact parameter ``sweet spot" for atmospheric characterization, independent of the sign of the active-region temperature contrast, whose location depends on the distribution of active regions. The distribution may be revealed by transit residuals as multiple planets probe different latitudes, while longitudes are sampled in time, such that the variance and frequency of the correlated scatter could constrain active-region filling factors, sizes, and separations. 

\end{abstract}


\section{Introduction} \label{sec:intro}

The characterization of transiting exoplanet atmospheres is limited by our understanding of their host stars. Stellar-surface heterogeneities introduce spurious spectral features in transmission spectra \cite[e.g.,][]{Pont2008, Sing2011} due to differences between the average intensity spectrum of the star and the chord occulted by the planet, known as the transit light source (TLS) effect \citep{Rackham2018, Rackham2019, RackhamEspinoza2023}. The TLS effect can mask or mimic possible evidence for planetary atmospheres, introducing features as large as 400 ppm for an Earth-like planet around an M dwarf star \citep{Rackham2018}. Observations with JWST have already proven stellar activity to be a limiting factor in atmospheric characterization of transiting small planets, from rocky worlds to sub-Neptunes \citep[e.g.,][]{Lim2023, Moran2023, Ahrer2025, Espinoza2025, Glidden2025, Radica2025}. 

The TLS effect depends on the distribution of active regions on the stellar surface, which is largely observationally unconstrained. The Sun is currently the only dwarf star with a fully resolved surface. Its magnetic features occupy two activity belts:  $\sim5^\circ$ to 30$^\circ$ in latitude ($b=0.087-0.500$) for spots and 5$^\circ$ to 40$^\circ$ ($b=0.087-0.643$) for faculae. For rapidly rotating stars, observations using Doppler imaging (DI) and Zeeman--Doppler imaging (ZDI) suggest a markedly different picture with magnetic features concentrated at high latitudes or even at the poles \citep{Strassmeier1990, Strassmeier2002, Strassmeier2009, Donati1992, Berdyugina2005}. 

The contrast between the solar active-region distribution and that of fast rotators can be understood from the physics of how magnetic flux reaches the surface. In stars with a radiative core, the magnetic field is generated by the action of the dynamo near the tachocline \citep{Charbonneau2020}. It is organized into discrete flux tubes anchored at the base of the convection zone with a free end that becomes buoyantly unstable and rises to emerge as active regions \citep{Fan2021}. The trajectory of magnetic-flux tubes through the convection zone is shaped by two forces: buoyancy, which drives the flux tube radially outward, and the Coriolis force, which deflects the rising tube poleward. On fast rotators, the Coriolis force dominates over buoyancy so rising tubes are deflected to high latitudes before they reach the surface, and flux emerges in polar or near-polar regions rather than in low-latitude belts \citep{Schuessler1992, Schuessler1996, DeLuca1997}. The latitude of flux-tube emergence, and thus active-region appearance, is therefore set by the balance between the two forces, and it shifts poleward with increasing rotation rate and depth of the convection zone. 

The Coriolis-force-mediated active-region distribution leads to a quieter equatorial belt, reminiscent of the Earth's doldrums, though not one fully devoid of magnetic activity \citep{Isik2024}. Planets that transit closer to the equator of a star with a strong Coriolis effect sample a transit chord that may be less representative of the average spectrum of the disk than those at moderate $b$, where the impact parameter ($b$) is the projected distance between center of the star and the planet at midtransit \citep[see e.g., Fig.~2 of][]{Winn2010}. Thus, the amplitude of the TLS effect may decrease with increasing impact parameter, presenting a testable hypothesis and a new dimension to consider for planning atmospheric-characterization observations. 

Multiplanet transiting systems offer an opportunity to measure how the wavelength-dependent TLS amplitude varies with impact parameter and constrain the latitudinal distribution of magnetic features on the stellar surface. The TRAPPIST-1 system of seven terrestrial planets straddling the habitable zone provides a unique testbed. The M8V host star, TRAPPIST-1, is highly magnetically active, but too faint ($m_V=17.02\pm0.20$) and slowly rotating \citep[$P_{rot}=3.30\pm0.14$,][]{Luger2017} to be accessible to DI or ZDI. Transmission observations of the inner TRAPPIST-1 planets (b, c, d, e) with JWST have yet to reveal clear evidence for atmospheric gases \citep{Lim2023, Espinoza2025, Glidden2025, Piaulet-Ghorayeb2025, Radica2025, Rathcke2025, Allen2026}. This is due in part to the low number of (published) observed transits (2--4), but also to the significant challenge of disentangling the planetary and stellar contributions to the signal. For the outer planets, 2-transit NIRSpec/PRISM transmission observations (0.6–\qty{5.3}{\um}) of planets g and h have been collected, but remain unpublished, while 5-transit NIRISS/SOSS observations (0.6–\qty{2.8}{\um}) of planet f were very recently released and showed no evidence for contamination by unocculted active regions \citep{Lim2026arXiv}.

While we invoke the Coriolis force as a physical mechanism to explain why active regions may concentrate to higher latitudes, we do not model the mechanism from first principles. Instead, we leave the specific functional form of the active-region distribution and its dependence on rotation rate as a generic parameterized form. We defer a physical derivation to later work, allowing our conclusions to be largely mechanism independent. Additionally, while the $\alpha$-$\Omega$ dynamo, driven by the twisting action of helical convection (the $\alpha$ effect) and the shearing action of differential rotation ($\Omega$ effect), is well motivated for stars with radiative cores, TRAPPIST-1 is fully convective and thus lacks a tachocline to anchor flux tubes. Fully convective stars likely operate via a different mechanism, such as a distributed dynamo \citep{Yadav2015}. A distributed dynamo would lead to active regions with properties similar to those of stars with only a convective layer. As M dwarfs have large convective turnover times \citep{Wright2018}, they have low Rossby numbers $(P_{\mathrm{rot}}/\tau_{\mathrm{conv}})$ compared with G dwarfs at the same rotation rate. As a low Rossby number indicates that rotation dominates over convection, this supports the concentration of active regions towards higher latitudes than in Sun-like stars.

In this Letter, we investigate how the wavelength-dependent amplitude of stellar spectral contamination for different planets in the system may depend on the latitudinal distribution of TRAPPIST-1's magnetic features and how it varies from planet to planet. Our Letter is organized as follows. In Section \ref{sec:methods}, we describe our minimal TLS effect model. Then, in Section \ref{sec:results}, we test distributions of magnetic activity that can cause the moderate-$b$ TRAPPIST-1 planets to have flatter transmission spectra than the low-$b$ planets. Finally, in Section \ref{sec:summary}, we summarize our results and their implications for future observations with JWST and other observatories. Appendix \ref{sec:appendixA} presents analytical solutions for the impact parameters that minimize the TLS effect given more general latitudinal distributions of magnetic activity.


\section{Methods} \label{sec:methods}

For our testbed, we use the TRAPPIST-1 system of seven Earth-sized planets (R$_{p}=0.755-1.129$ R$_{\oplus}$; \citealt{Agol2021}). The TRAPPIST-1 planets span a range of impact parameters from $b=0.0950^{+0.0650}_{-0.0610}$ (planet b) to $b=0.379^{+0.018}_{-0.014}$ (planet g) \citep{Agol2021}. We note that the sequence of planets with respect to their impact parameters is not the same as their sequence in orbital distance. We assume a stellar radius of R$_{*}=0.1192\pm0.0013$ R$_{\odot}$ \citep{Agol2021}. The photometric variability of TRAPPIST-1 suggests that a network-like distribution of small-scale magnetic features is dispersed across the stellar surface rather than concentrated in a few large areas \citep{Rathcke2025}. To represent this, we parameterize the active regions by their covering fraction without specifying the number or size of individual features. 

We use a minimal model to explore the effect of different spatial distributions of active regions. As we do not have a functional relationship between the Coriolis force and the spatial distribution of stellar heterogeneity, we first invoke a simple two-zone model to demonstrate how a given distribution shapes variations in the TLS effect with $b$. We divide our stellar surface into a polar region and an equatorial region, defined by a latitudinal boundary, and assume hemispheric symmetry. The boundary is set at a latitude of $\pm15^\circ$ ($b=0.259$), selected to divide the TRAPPIST-1 planets in two groups, inner planets (b, c, d, and e) and outer planets (f, g, and h). The specific results shown here are a consequence of our choice of this representative zonal boundary. We further evaluate alternative choices for the distribution of active regions in Appendix \ref{sec:appendixA} to generalize our findings and demonstrate possible analytical solutions for the impact-parameter sweet spot.

We use 1D disk-integrated spectra from {\tt PHOENIX} \citep{Hauschildt1997, Husser2013} for both the active and quiet stellar components. As our results are far more sensitive to the adopted distribution of magnetic features than the choice of spectral models, the classical {\tt PHOENIX} grids are sufficient here. Higher-fidelity, 3D radiative-MHD grids offer no practical improvement given our other assumptions. Fluxes are calculated from a linear combination of the photospheric flux and active-region flux, weighted by the projected area of the star and planet and the relevant covering fraction. We calculate the transit depth ($R_p^{2}/R_s^{2}$) as the square of ratio of the area occulted by the planet compared with the star. For simplicity, we do not further add in the effect of how limb darkening changes with wavelength nor impact parameter. This is a reasonable approximation for the chord-averaged occulted flux for transit chords relatively close to the disk center, such as those in the TRAPPIST-1 system (b $\le$ 0.38). Additionally, we assume that planetary orbits are aligned with the stellar-rotation axis given the compact, coplanar, resonant nature of the TRAPPIST-1 system \citep{Gillon2017} and the low projected stellar obliquity \citep{Hirano2020, Brady2023}.

We compare our models to transmission observations of the TRAPPIST-1 planets using reduced single-transit observations of JWST NIRSpec/PRISM for planets b \citep{Rathcke2025}, c \citep{Rathcke2025}, d \citep{Piaulet-Ghorayeb2025}, and e \citep{Espinoza2025}.

\section{Results} \label{sec:results}

We demonstrate that the latitude-dependent spot distributions may have observational implications for transiting planets. Here, we evaluate the impact for a range of active-region distributions and brightness contrasts, which change with temperature, using our fiducial two-zone model. 

\subsection{Moderate-Impact-Parameter Planets May Exhibit A Reduced TLS Effect} 

\begin{figure*}[!ht]
    \centering
    \includegraphics[width=\textwidth]{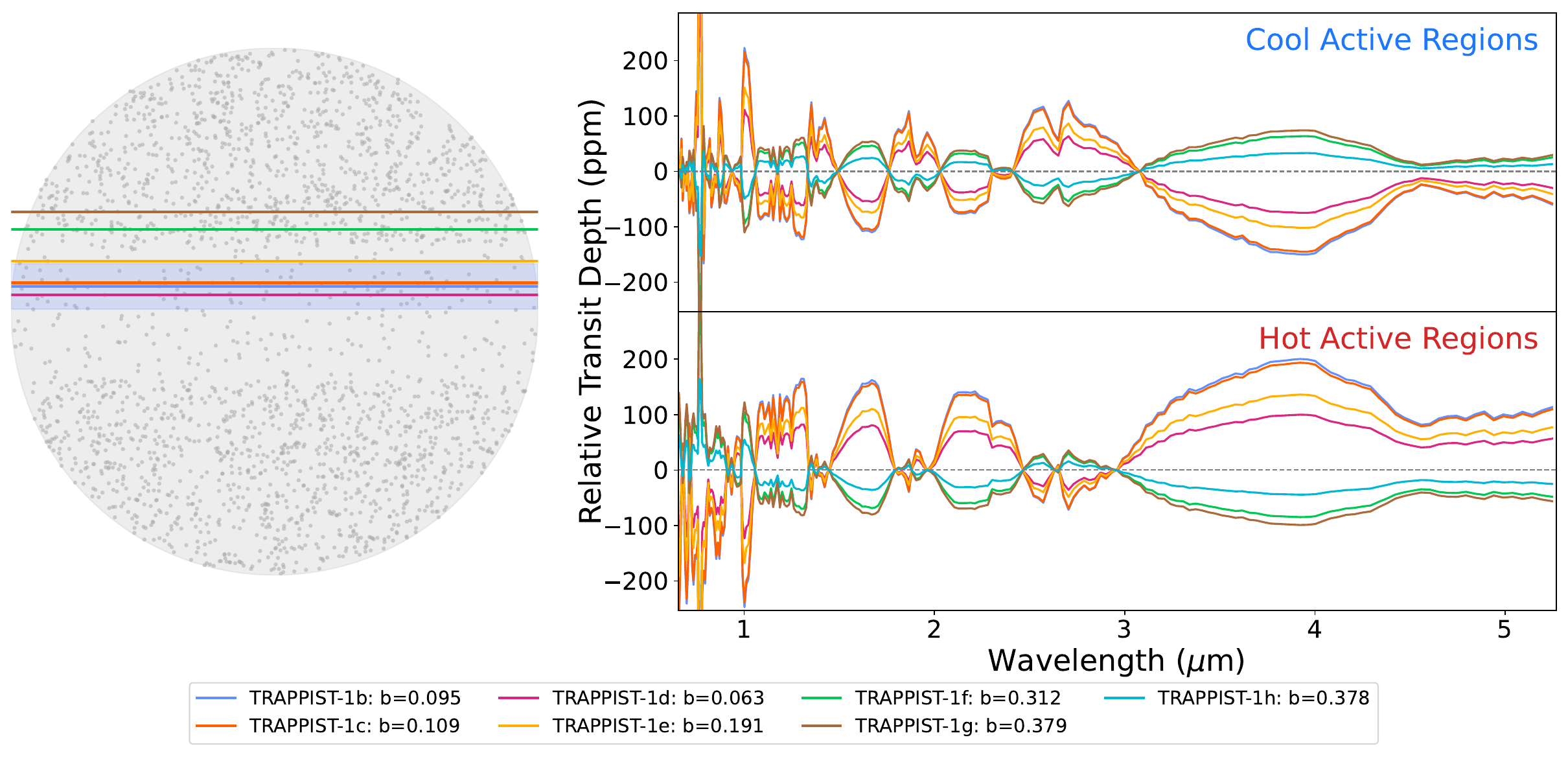}
    \caption{Differences in the TLS effect for the TRAPPIST-1 planets for our minimal two-zone model. The left panel shows a representative diagram of the stellar surface with increased spottedness in the polar regions. Lines are drawn to represent the center of the transit chord for each of the TRAPPIST-1 planets. As a representative example, the width of planet b's transit chord is shaded in blue. In the right panel, the relative transit depth is shown assuming only stellar heterogeneity made up of cooler regions (upper) and hotter regions (lower) with no contribution from a planetary atmosphere over the wavelength range relevant to JWST NIRSpec/PRISM. The higher-impact-parameter planets (f, g, and h) are less impacted by the TLS effect than the lower-impact-parameter planets given our choice of zonal boundary.} 
    \label{fig:spoteffect}
\end{figure*}

The main result from our analysis is that planets at moderate impact parameter ($b\approx0.4-0.6$ for the example distributions shown in Appendix \ref{sec:appendixA}) may fall within an observational sweet spot for the TLS effect. This is because their occulted region may be more similar to the disk-averaged stellar spectrum than those at low impact parameters if the equatorial belt is relatively devoid of active regions. The location of the sweet spot depends on the distribution of active regions.

We first show this effect for a two-zone stellar surface with covering fractions ($f$) for the pole and equator of $f_{pole}=0.4$, $f_{equ}=0.1$; temperature contrast between the photosphere (T$_{phot}$=2566 K, $\log g = 5.5~\mathrm{cm\,s^{-2}}$) and active regions, $\Delta T=\pm100$~K; and a 15$^\circ$ latitude boundary. Figure \ref{fig:spoteffect} shows a schematic diagram of our generated stellar surface and transit latitudes for the TRAPPIST-1 planets on the left. Individual spots are shown for illustrative purposes only as our model does not represent discrete active regions, but rather uses the filling factors to weight a linear combination of the photosphere and active-region spectral flux. On the right, the upper panel shows the modulation in transit depth due to the TLS effect for each planet assuming active regions cooler than the photosphere while the lower panel is for hotter active regions. For the moderate-$b$ planets (f, g, and h), the maximum amplitude of the features from the heterogeneity is $\sim30$--$80$\,ppm compared with $\sim70$--$150$\,ppm for the low-$b$ planets (b, c, d, and e). Additionally, the sign of the features is inverted between the low-$b$ and moderate-$b$ planets with the low-$b$ planets showing negative contamination most prominently in the broad feature centered around $\sim$\qty{4}{\um}. 

Transmission spectral contamination is wavelength dependent and most prominent where the star's spectrum is rich with features from molecules like TiO, VO, FeH, and H$_2$O, and atomic lines such as K I and Na I. The water bands in the stellar atmosphere, in particular, strongly shape the spectral features. The strength of contamination is also time variable due to stellar rotation and the evolution of active regions. However, for simplicity, our model does not include time variability.

\begin{figure*}[!ht]
    \centering
    \includegraphics[width=0.85\textwidth]{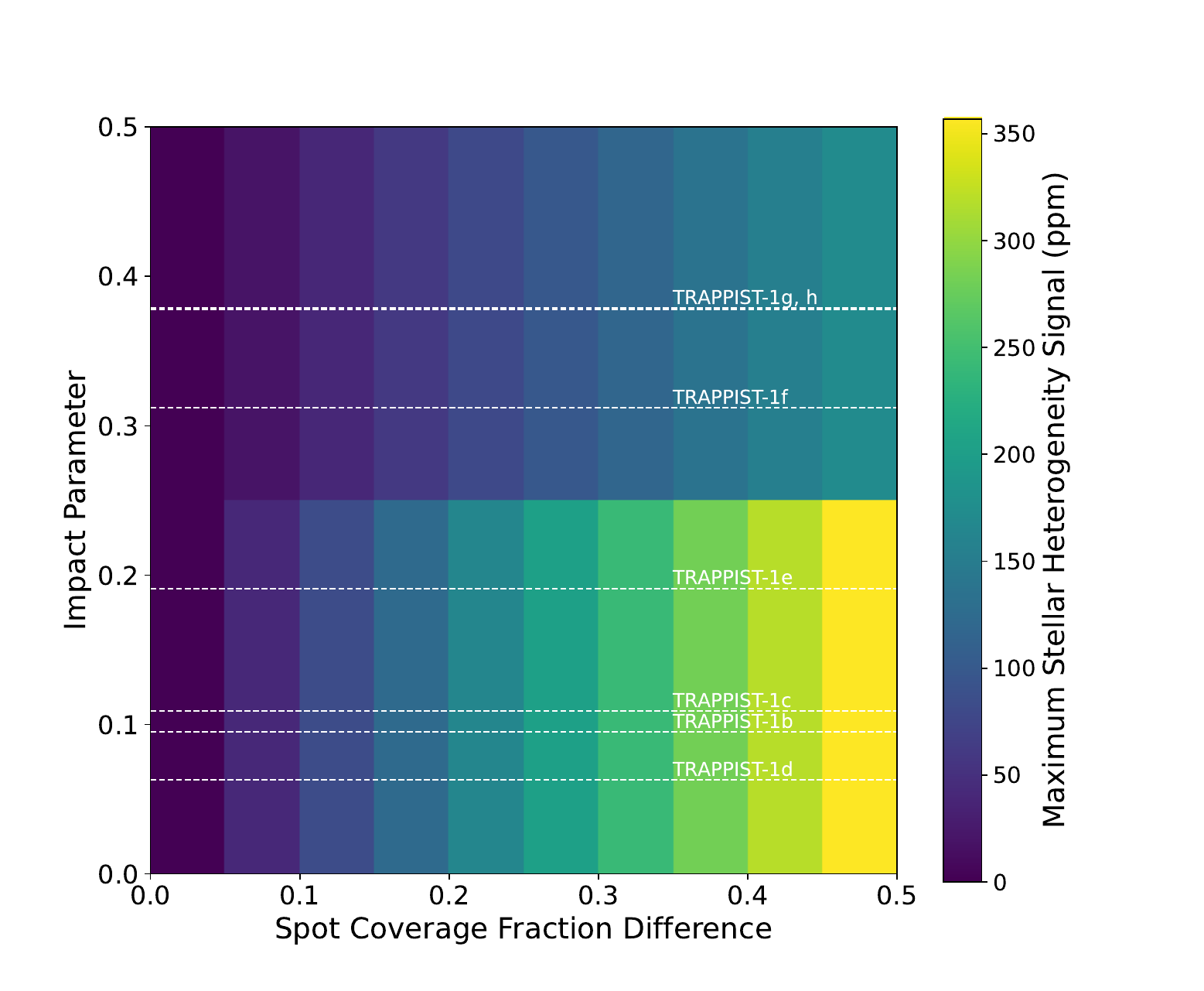}
    \caption{The maximum peak-to-peak amplitude of the TLS effect over a range of impact parameters and spot-coverage fractional differences between the polar and equatorial regions for our fiducial two-zone model. As a representative example, we assume cold spots ($\Delta T=-100$\,K) and an equatorial band up to 15$^\circ$. We fix $f_{pole}$ to 0.5 and allow $f_{eq}$ to range from 0 to 0.5. The inner (low-$b$) planets demonstrate up to ${\sim}2\times$ the TLS as the outer planets.}
    \label{fig:spotgrid}
\end{figure*}

Preserving our two-zone model, we assess our findings over a range of covering fractions and show the resultant maximum peak-to-peak amplitude within the JWST NIRSpec/PRISM bandpass \citep{Jakobsen2022} over a range of impact parameters (Figure \ref{fig:spotgrid}). We hold $f_{pole}$ at 0.5 and vary $f_{equ}$ from 0 to 0.5, keeping our zonal boundary at 15$^\circ$ as before, approximately midway between the inner and outer TRAPPIST-1 planets. We adopt the average TRAPPIST-1 planet radius as the planets are similar in size. With these distributions, the outer planets have a reduced TLS effect and the impact of the TLS effect increases with the spot-coverage fraction difference, leading to features that would swamp those anticipated from a planetary secondary atmosphere.

\subsection{Preferred Moderate-\lowercase{$b$} Transit Geometry is Robust to Active-region Temperature Contrast}

Our finding of a weaker TLS effect for the outer TRAPPIST-1 planets for our fiducial two-zone distribution is independent of the sign of the temperature contrast between the active regions and photosphere. We demonstrate the wavelength-dependent effect with representative hot regions of $\Delta T=+100$~K with $f_{pole}=0.4$, $f_{equ}=0.1$, and 15$^\circ$ latitude boundary (Figure \ref{fig:spoteffect}; lower right). As with the cool spots, the amplitude of the features is smaller for the outer planets compared with the inner planets. However, the sign of the features has been inverted.

The nature of `hot spots' on TRAPPIST-1 is currently unclear. Small bright features have been proposed as the source of rotational variability \citep{Morris2018a}. On the Sun, bright features associated with small-scale concentrations of magnetic features are readily observed. However, 3D MHD simulations point to an opposite effect on cool red dwarfs where small-scale concentrations lead to the formation of dark features \citep{Kostogryz2026arXiv, Shapiro2026arXiv}. Bright features could be caused by a not yet identified physical mechanism, which may be connected to strong flaring activity. Alternatively, it may be simply quiet regions between cold magnetic spots and faculae. 

\subsection{Intrinsic Degeneracies between Active-region Temperature and Distribution}

\begin{figure*}[!ht]
    \centering
    \includegraphics[width=\textwidth]{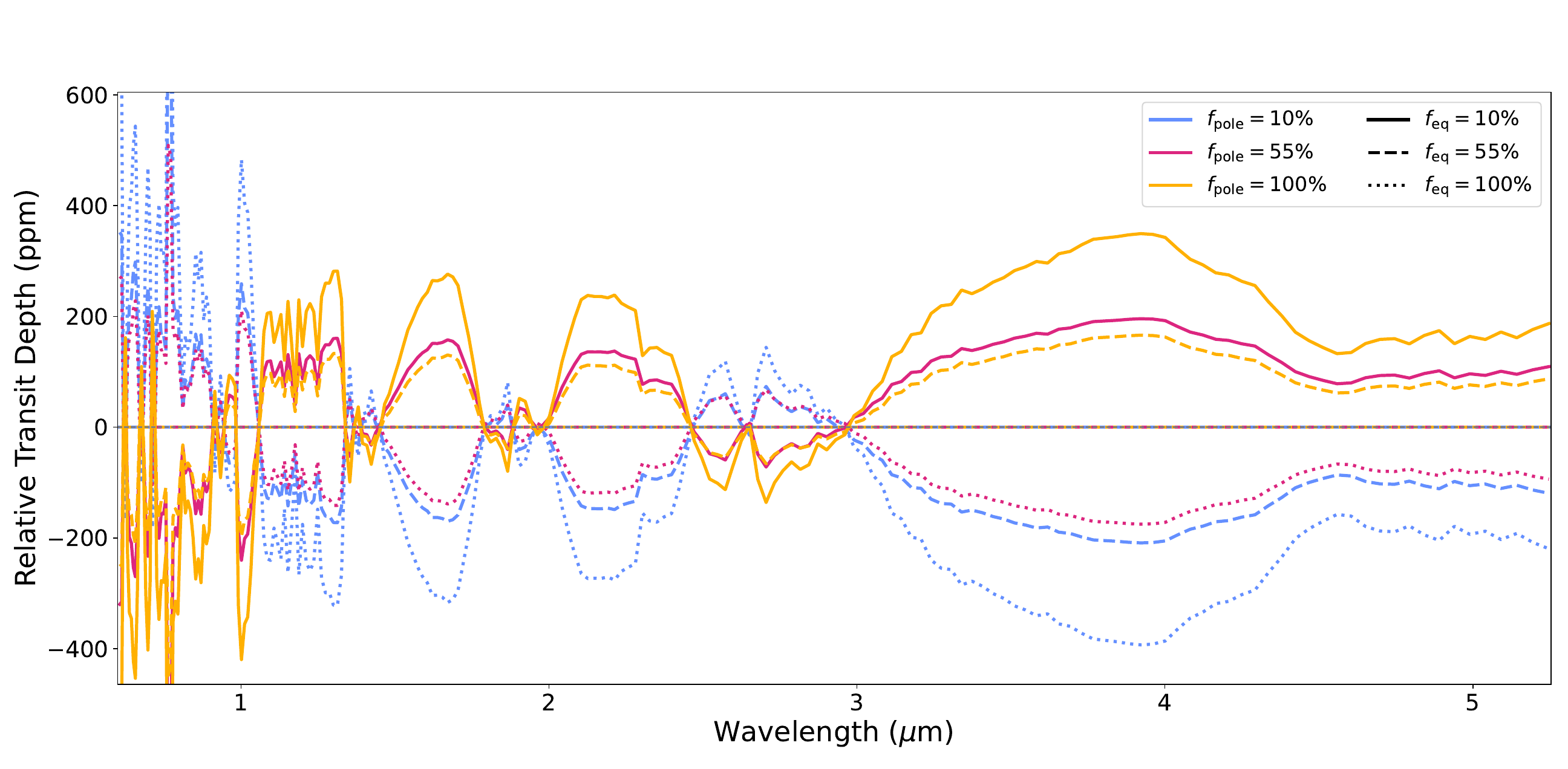}
    \caption{Degeneracy between filling-factor differences between the polar and equatorial regions and the temperature contrast (Figure \ref{fig:spoteffect}). Each curve shows the error in transit depth due to the TLS effect for an assumed polar filling factor ($f_{pole}$; line color) and equatorial filling factor ($f_{eq}$; line style). For example, the solid yellow line has 100\% filling factor at the poles (yellow) and 10\% filling factor at the equator (solid line). This degeneracy makes it challenging to fit simple temperature and filling-factor models with observed data.}
    \label{fig:degeneracy}
\end{figure*}

The temperature contrast cannot uniquely change the peak-to-peak amplitude and sign of the features. Differences in the spot-covering fraction between the equatorial and polar regions are degenerate with the temperature contrast \citep{Strassmeier2009, Luger2021}. Thus, the TLS effect depends only on the product of the active-region covering fractions and intensity contrast. In Appendix~\ref{sec:appendixA} we give an analytical solution for this dependence for a simple active region distribution model. By comparing the results in Figure \ref{fig:spoteffect} and Figure \ref{fig:degeneracy}, we show this degeneracy using TRAPPIST-1~e as an example.  

When unocculted, cool spots artificially increase the transit depth \citep[e.g.,][]{Pont2008,Czesla2009, Sing2011}. In contrast, when they are occulted, the transit depth decreases \citep[e.g.,][]{Silva2003,Pont2007,Bruno2018}. In comparison, hot regions have the opposite effect. Both types of contamination may occur at the same time, leading to a mixture of both positive and negative contamination. 

It is underappreciated that cooler (i.e., darker) regions can produce both positive or negative contamination based on the distribution of the active regions. If the chord is less spotted than the disk, it can lead to positive contamination, but if the chord is more spotted than the disk, it can cause negative contamination without the need to invoke hot regions. For a given contamination, this implies a degeneracy between the brightness of the magnetic feature(s) and their spatial distribution on the stellar surface. By invoking the expected distribution of stellar spots emerging towards the poles as mediated by the Coriolis effect, we can break this degeneracy for cold spots. Thus, the appearance of negative contamination in a transit spectrum alone is not evidence for bright features as it is degenerate with the transit chord being more heavily covered by cool spots than the full stellar disk.

\subsection{Testing Predictions Against Inner-planet TRAPPIST-1 Transit Observations}

\begin{figure*}[!ht]
    \centering
    \includegraphics[width=\textwidth]{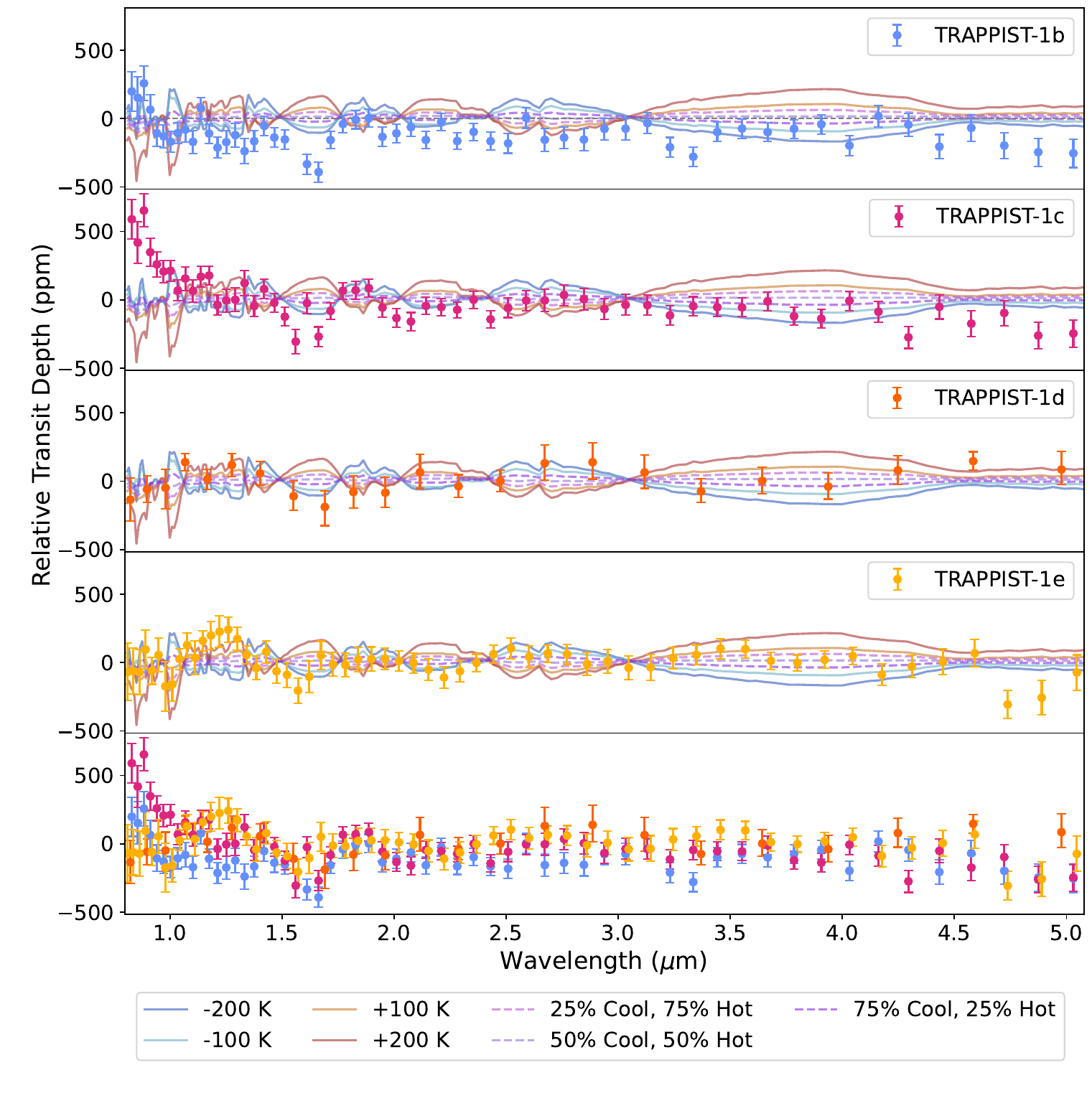}
    \caption{Planet-to-planet comparison of observed single-transit low-$b$ planet spectra with TLS-effect models. Transit depth in parts-per-million (ppm) is shown on the y-axis and wavelength is shown in microns on the x-axis. The top four panels show JWST NIRSpec/PRISM data for the inner planets individually compared with a range of expected contamination models while the lowest panel shows the planetary spectra compared with each other. We show models using cool active regions, hot active regions, and mixtures of hot and cold active regions. While the minimal models alone cannot provide a fit to the data that would allow for TLS mitigation, they can inform on planet-to-planet trends.}  
    \label{fig:innerplanets}
\end{figure*}

\begin{figure*}[!ht]
    \centering
    \includegraphics[width=\textwidth]{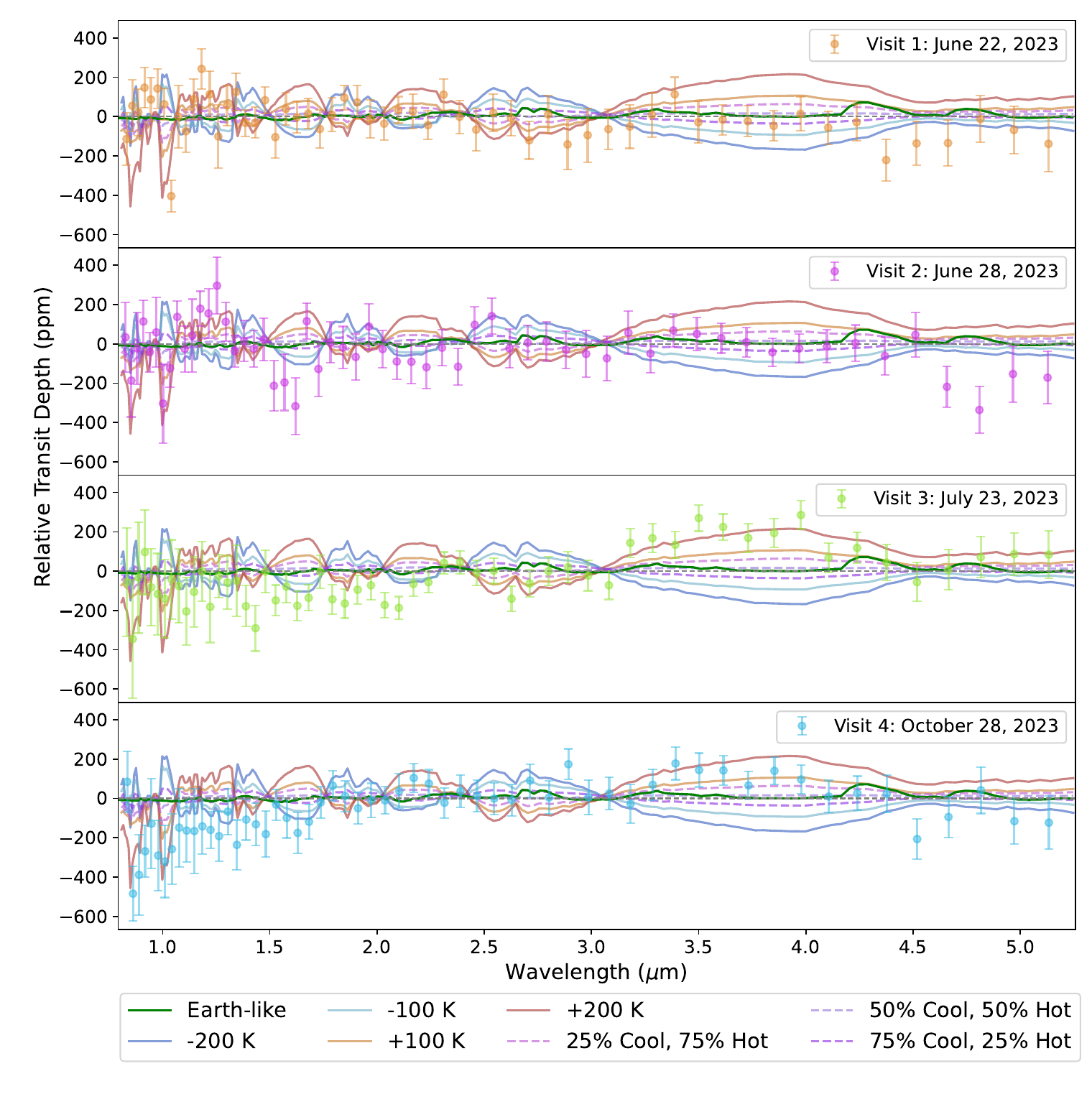}
    \caption{Visit-to-visit variability of TRAPPIST-1~e transmission spectra compared with TLS-effect models for a hot, cold, and mixed active region along with a representative Earth-like terrestrial atmosphere model (dark green). While none of the models fit the data across all wavelengths, some trends can be seen. For example, Visit 3, which has a midtransit flare, shows broadly more agreement with the hotter models compared with the quieter first two visits, in particular between 3 and \qty{4.5}{\um}.} 
    \label{fig:t1efourvisits}
\end{figure*}

The degeneracy mentioned above makes it challenging to map active-region temperatures and filling factors to observed transmission spectra. To demonstrate this, we compare a set of models against both planet-to-planet variations (Figure \ref{fig:innerplanets}) and visit-to-visit variations (Figure \ref{fig:t1efourvisits}). We show each transmission spectrum separately and compare them with cool, hot, and three-component (cool, hot, and photosphere) models, assuming $f_{pole}=0.4$, $f_{equ}=0.1$. Significant variability between observations is expected given the active nature and rotation of TRAPPIST-1. If regions with more activity evolve on shorter timescales than those with less, then increasing stellar heterogeneity towards the poles may further imply that while low-$b$ planets are anticipated to have an increased TLS effect, the variability of that effect may be less than for planets at moderate $b$. This is because low-$b$ planets occult a relatively quiet chord that may be more temporally stable than the more active chord occulted by the moderate-$b$ planets. Additional observations from the JWST TRAPPIST-1 e/b Program \citep[JWST-GO \#6456 and \#9256, PI N. Allen \& N. Espinoza;][]{Allen2026} will be able to test this prediction with a larger number of transits. 

In Figure \ref{fig:innerplanets}, we use the data for the first transit observed with JWST NIRSpec/PRISM for planets b \citep{Rathcke2025}, c \citep{Rathcke2025}, d \citep{Piaulet-Ghorayeb2025}, and e \citep{Espinoza2025}. For a single transit, contribution from a planetary secondary atmosphere (if any) would be negligible compared with the stellar and instrumental noise. Each planet transmission spectrum shows a significant TLS effect, which is not consistent between the planets. Additionally, we caution that differences in data reduction methodologies complicate rigorous comparison of the effect between planets, especially at shorter wavelengths ($<$\qty{2}{\um}) \citep[see, e.g.,][]{Espinoza2025, Allen2026}.

In Figure \ref{fig:t1efourvisits}, we show the variability between each of the four-transit observations of TRAPPIST-1~e \citep[JWST-GTO 1331, PI N. Lewis;][]{Espinoza2025}, which complicated secondary-atmosphere characterization \citep{Glidden2025}. The Earth-like atmospheric model from \citet{Glidden2025} is shown in green for comparison with the TLS-effect models. As reported in \cite{Espinoza2025}, there is no single model of stellar heterogeneity that adequately fits the data. However, the last two visits, which show increased stellar activity, are more consistent with a hot model than the quieter (flatter) earlier two visits. In particular, Visit 3, which has a midtransit flare, shows the most evidence for hot regions.

\section{Summary and Discussion} \label{sec:summary}

Transmission spectra present our best opportunity to detect and characterize the atmospheres of temperate rocky worlds with current and near-term observatories. However, the TLS effect remains a significant barrier. We must be able to precisely extract the planetary contribution to an observation ($\sim10$ ppm). As the true Coriolis-force-mediated active-region spatial distribution is unknown, we used a minimal two-zone model to demonstrate the effect of different temperature contrasts and active-region distributions on stellar contamination for a range of impact parameters. In Appendix \ref{sec:appendixA}, we generalize to other distributions. We use the TRAPPIST-1 system as a proving ground, though our results also apply to other systems where the host star rotates sufficiently rapidly to concentrate active regions towards the pole. Our primary conclusions are as follows:

\begin{enumerate}
    \item The TLS effect may be mitigated by observing planets at moderate impact parameter if they sample heterogeneities more representative of the disk-averaged spectrum. This occurs for host stars with a polar-enhanced active-region distribution. The exact location of the sweet spot depends on the details of the active-region distribution ($b\approx0.4-0.6$ for the cases in Appendix \ref{sec:appendixA}). This will be empirically tested through multiplanet systems such as TRAPPIST-1.

    \item A gradient of increasing active regions with latitude may lead to lower-impact-parameter planets having greater contamination, but less variability in their contamination on a visit-to-visit basis, though this depends on the specifics of the active-region distribution and evolution timescale. The inverse is also predicted. This effect can also be empirically tested.
    
    \item The TLS effect may lead to more degenerate spectral contamination than previously appreciated as cool features can cause either positive and negative contamination, depending on the comparative distribution between occulted and unocculted regions. For example, cool active regions can produce negative contamination, mimicking faculae, if the transit chord is preferentially spotted compared with the disk-integrated spectrum.
 
\end{enumerate}

We now assess the implications of our findings, limitations, and future work. 

\subsection{Implications for Stellar-Contamination Correction Using Back-to-back Transits}

Observations of back-to-back transits have been considered as a way to mitigate stellar contamination without stellar spectral models \citep[e.g.,][]{Allen2024jwstprop, Allen2025jwstprop, Rathcke2025, Allen2026}. If two planets at similar impact parameters transit close in time, and one is airless, dividing their spectra removes the imprint from the star. \citet{Rathcke2025} used the technique on a single transit each of TRAPPIST-1~b and c taken with JWST NIRSpec/PRISM, finding a 2.5$\times$ reduction in the stellar noise for wavelengths below \qty{2}{\um}. As the transits were partially coincident with planet c's egress occurring at the same time as planet b's ingress, stellar-surface evolution should be negligible, allowing for an optimal test of the back-to-back technique. Despite the favorable geometry and timing, the technique fell short of removing the stellar imprint. In particular, a strong slope at shorter wavelengths ($\sim$\qtyrange{0.8}{1.5}{\um}) persisted. The on-going 15-transit program for TRAPPIST-1~b and e \citep[GO \#6456 and \#9256, PI Allen \& Espinoza,][]{Allen2024jwstprop, Allen2025jwstprop} will further test this method.

While planets b and c have a $\sim85$\% overlap between the union of their transit chords, the overlap between planets b and e drops to only $\sim24$\%. However, it is important to note that the overlap is poorly constrained, driven largely by the uncertainty on planet b's impact parameter of $b=0.0950^{+0.0650}_{-0.0610}$ \citep{Agol2021}. From the first three visits of the program, \citet{Allen2026} demonstrated that TRAPPIST-1's frequent flaring can cause the star's spectrum to change within the time frame of a single visit. Increasing active regions with latitude may pose additional challenges for this technique.  If there is a gradient in increasing active regions with latitude, TRAPPIST-1~b likely samples a quieter transit chord compared with planet e, and may show more of the TLS effect than e. However, variations in the TLS effect may be be less in planet b than in e on a visit-to-visit basis if regions more rich in active regions change more with time.

\subsection{Mapping Stellar Activity in Latitude and Longitude} 

We can constrain stellar-activity latitudinal distributions from multiplanet transit observations and test predictions of a Coriolis-force-mediated distribution. Spectral profiles of occulted features \citep[e.g.,][]{Sing2011, Murray2026} can help constrain their nature while impact parameter can inform on their latitudinal distribution. A latitude-dependent active-region distribution should produce impact-parameter-dependent in-transit correlated scatter. Measuring the variance, autocorrelation, and frequency distribution of transit residuals across multiple TRAPPIST-1 planets offers a direct way to test the proposed latitudinal structure and to constrain the characteristic sizes and separations of the underlying active regions. Additionally, time-resolved transit spectra can be used to unlock the longitudinal distribution. As it transits, the planet would act as a coronagraph-like probe, where the planetary radius sets the smallest features it can resolve. Complementary techniques, such as those proposed by \citet{Morris2018b}, can aid in verifying the distributions. These indirect approaches provide alternative mechanisms to map stellar activity for systems which cannot be reached by current ZDI measurements.

\subsubsection{Towards a Self-consistent Model-observation Framework} 

The interpretation of terrestrial exoplanet atmospheres is complicated by the lack of high-fidelity synthetic stellar spectral models which accurately include magnetic effects. Differences between observed spectral profiles and modeled spectra can help to improve the stellar spectral models. Through an iterative process between observations of occulted magnetic features and enhanced 3D radiative-magnetohydrodynamic (MHD) models, we can move closer to both understanding the physics of cool stars and mitigating stellar contamination. 

As an example, planets around stars like TRAPPIST-1 occult not only network-like active regions, but also flares. Flares are one source of contamination that is particularly challenging to mitigate, which we do not account for in our minimal model. Unlike spots or faculae which exhibit a static spectral profile but time-dependent surface coverage, flares are stochastic. Their spectral energy profile evolves dynamically in time \citep{Howard2023, Howard2025} and occurs on timescales of JWST transit observations (a few hours) to minutes \citep{Vasilyev2026}. 

The \citet{Vasilyev2026} power-law flare frequency distribution indicates that flares with energies of $10^{29}$ erg occur approximately ten times per day on TRAPPIST-1. Flares with the lowest-detectable energies are the most common, and flare frequency is expected to increase at smaller energies than we can detect with JWST spectrophotometry. \citet{Morris2018a} propose that the strongest flares occur during smaller-amplitude brightening on a longer timescale. Recent HST/STIS and VLT observations also report variability induced by microflares on a sub-hour timescale, producing a residual activity floor \citep{Berardo2026}. While K2 observations show no correlation between flare occurrence and rotational phase \citep{Vida2017}, multitransit observations provide the opportunity to improve constraints on the surface distribution of flares and their associated active regions, potentially allowing for improved mitigation strategies.

\subsubsection{Impact-parameter-dependent Red Noise as a Test of Latitudinal Active-region Structure}

A latitude-dependent active-region distribution should affect not only the mean TLS contamination spectrum, but also the time-domain structure of the transit light curve. In our minimal model, each transit chord is represented by an average active-region covering fraction. In reality, however, the occulted chord samples a finite and spatially structured distribution of spots, faculae, and other magnetic-network elements. As the planet moves across this structure, spatial inhomogeneities along the chord are mapped into time-correlated residuals in the transit light curve.

This provides an additional observational test of the latitude-gradient hypothesis. If active-region coverage or morphology changes with latitude, then planets with different impact parameters should exhibit different in-transit residual covariance structures after removal of the best-fit transit model and common-mode systematics. The relevant diagnostics include the excess in-transit variance relative to the out-of-transit baseline, the residual autocorrelation time, and the power spectral density of the in-transit residuals. These quantities should depend on the local active-region filling factor, brightness contrast, characteristic feature size, and mean spacing along the occulted chord.

A simple scaling illustrates the potential observability of this effect. A magnetic feature of characteristic radius ($r_{spot}$) produces a residual timescale set by the projected crossing time of the planet over that feature ($T_{\mathrm{tr}}$), approximately $t \sim \tfrac{(R_{\mathrm{p}} + r_{\mathrm{spot}})\, T_{\mathrm{tr}}}{R_{\star}}$ for low-impact-parameter transits, with a weak additional dependence on $b$ through the chord length. For TRAPPIST-1-like planets with $R_{\mathrm{p}}/R_{\star} \approx 0.085$ and transit durations of order tens of minutes, spots with radii of $\sim 0.03\text{--}0.05\,R_{\star}$ produce correlated residuals on timescales of a few minutes, within the range probed by JWST time-series observations. The amplitude of the residuals depends on the feature contrast and filling-factor fluctuations within the planet’s projected disk. Multiple transits can therefore be used not only to measure the mean TLS spectrum, but also to constrain the spatial granularity of the stellar surface.

This test is complementary to the planet-to-planet comparison of transmission spectra. Two active-region distributions with similar chord-averaged filling factors could produce similar TLS amplitudes but different residual power spectra if one consists of many small, well-mixed features and the other of fewer, larger structures. Conversely, the absence of impact-parameter-dependent in-transit red noise would favor either very small features below the JWST detection threshold, a nearly uniform distribution at the spatial scales sampled by the planets, or residual noise dominated by flares and instrumental systematics. Future multitransit JWST observations of TRAPPIST-1 planets spanning different impact parameters could therefore directly test whether the proposed latitudinal active-region gradient is present and provide an empirical benchmark to guide the developing theory of how rotation mediates the distribution of active regions for fully convective stars.

\begin{acknowledgments}
We thank the anonymous reviewer for their time. ERC Synergy grant: This work was funded by the European Research Council (ERC) under the European Union’s Horizon 2020 research and innovation program (grant no. 101118581) and JWST-GO \#3593 (PI: S. Seager). The observational data used within this Letter can be accessed through the Mikulski Archive for Space Telescopes (MAST) at the Space Telescope Science Institute via doi 10.17909/4zcj-k904 \citep[TRAPPIST-1~b and c;][]{Rathcke2025}, doi 10.17909/2jzw-7m72 \citep[TRAPPIST-1~d;][]{Piaulet-Ghorayeb2025}, and doi 10.17909/yzwd-vq54 \citep[TRAPPIST-1~e;][]{Espinoza2025}. This paper reports work done in collaboration with the JWST Telescope Scientist Team (PI: M. Mountain; {\tt https://www.stsci.edu/$\sim$marel/jwsttelsciteam.html}). Funding is provided to the team by NASA through grant 80NSSC20K0586. A.G. acknowledges the MIT Office of Research Computing and Data for providing high performance computing resources that have contributed to the research results reported within this Letter. N.K. and A.I.S acknowledge support by the Volkswagen Foundation (grant 9E126). N.K. and V.V. acknowledge support by the German Aerospace Center (DLR) grants ``PLATO Data Center'' $50$OO$1501$ and $50$OP$1902$. This material is based upon work supported by the National Aeronautics and Space Administration under Agreement No.\ 80NSSC21K0593 for the program ``Alien Earths.'' The results reported herein benefited from collaborations and/or information exchange within NASA's Nexus for Exoplanet System Science (NExSS) research coordination network sponsored by NASA's Science Mission Directorate. 

\end{acknowledgments}


\appendix
\section{Analytic Solution to the Impact-parameter Sweet Spot}\label{sec:appendixA}

We can construct an analytical solution to the impact-parameter sweet spot (i.e., the one that minimizes the TLS effect), assuming that surface activity follows an arbitrary functional form of $f(b)$. While we have explored the effect of $f(b)$ increasing monotonically with $b$, consistent with the expectation that magnetic activity may increase towards the poles, driven by the Coriolis force, the formulae below are independent of a specific physical mechanism. We assume a photosphere temperature, $T_0$, and active-region temperature of $T_0 +\Delta T$. This sets the emergent spectral fluxes ($S_{\lambda}$) to be $S(T_0)$ and $S(T_0 + \Delta T)$, respectively, where we drop the $\lambda$ subscript for brevity. We further define $S_0 \equiv S(T_0)$ and $\Delta S \equiv S(T_0 + \Delta T) - S(T_0)$. While $\Delta S$ is generally wavelength dependent, $\Delta S$ drops out of the key equations we derive below (e.g., eqs.~[\ref{eq:bminPL}, \ref{eq:bmidzone}]), so that these are {\it not} wavelength dependent. We consider only $b\geq0$, assuming hemispheric symmetry. To isolate the effect of the active-region distribution on the sweet spot, we neglect other effects (e.g., limb darkening). For monotonically increasing $f(b)$, we expect limb darkening to shift the sweet spot towards lower $b$, though the exact behavior depends on both $f(b)$ and the choice of limb-darkening law.

The disk-average flux over the surface of the star is:
\begin{align}
\langle S \rangle 
&= \frac{4}{\pi} \int_{0}^{1} dy \int_{0}^{\sqrt{1-y^2}} dx \cdot S(x, y) \notag\\[6pt]
&= S_0 + \left( \frac{4}{\pi} \right) \Delta S \int_{0}^{1} dy\, f(y)\, (1 - y^2)^{1/2} \quad ,
\label{intavstar}
\end{align}
where we used that the average flux along a chord is:
\begin{equation}
\langle S \rangle_{\text{chord}}(b) = S_0 + \Delta S \cdot f(b)
\end{equation}
The TLS effect is minimized at the impact parameter $b_{\text{min}}$ for which $\langle S \rangle = \langle S \rangle_{\text{chord}}(b_{\text{min}})$, which implies:

\begin{equation}
f(b_{\text{min}}) = \left(\frac{4}{\pi}\right) \int_0^1 dy \, f(y)\,(1 - y^2)^{1/2}
\end{equation}

\bigskip

As an illustrative example, we can consider $f$ as a power-law (PL) of the form:
\begin{equation}
f_{\text{PL}}(y) = f_{\text{pole}} \cdot y^{\alpha} \quad .
\end{equation}
Here $f_{\text{pole}}$ is the active fraction at the pole ($y=1$), while the exponent $\alpha$ measures the extent to which the active regions are limited to the polar region (high $\alpha$) or extend over most of the stellar surface (low $\alpha$). Then
\begin{align}
f_{\text{PL}} (b_{\min,\,\text{PL}})
&=  \left(\frac{4 f_{\text{pole}}}{\pi}\right) \int_0^1 y^{\alpha} \,(1 - y^2)^{1/2}\, dy \notag\\
&= \left( \frac{2 f_{\text{pole}}}{\pi}\right) 
B \left(\frac{\alpha+1}{2},
\frac{3}{2} \right)
\end{align}
where $B$ is the beta function \citep{Gradshteyn1980}. Substitution of $f_{\text{PL}}$ on the left-hand side, and expression of the beta function in terms of Gamma functions on the right-hand side yields
\begin{equation}
b_{\text{min,PL}} = 
\left[ \frac{\Gamma\left(\frac{\alpha + 1}{2}\right)}{\sqrt{\pi} \, \Gamma\left(\frac{\alpha + 4}{2}\right)} \right]^{\frac{1}{\alpha}}
\label{eq:bminPL}
\end{equation}
This expression for $b_{\text{min,PL}}$ increases monotonically with $\alpha$. For $\alpha$ ranging from, e.g.,  $\frac{1}{5}$ to $5$, $b_{\text{min,PL}}$ ranges from $0.334$ to $0.627$.
Simple analytical results are obtained for certain integer values of $\alpha$, such as 
\begin{align}
b_{\min,\,\text{PL}} &=
\frac{4}{3\pi} \>\>\>\>\approx 0.424 
\qquad (\text{for } \alpha =1), \quad \\
b_{\min,\,\text{PL}} &=
\frac{1}{2} \quad\>\> = 0.500 \qquad (\text{for } \alpha =2), \quad \\
b_{\min,\,\text{PL}} &= 
2^{-3/4} \approx 0.595 \qquad 
(\text{for } \alpha =4), \quad 
\end{align}
corresponding respectively to linear, quadratic, and quartic functions $f_{\text{PL}}$.
The size of the TLS effect depends on
the value of $\langle S \rangle_{\text{chord}}(b) - \langle S \rangle$, which is zero for $b = b_{\min}$. Given that $f_{\text{PL}}(y)$ is a monotonically increasing function, the size of the TLS increases with $|b - b_{\min}|$ on either side of $b_{\min}$ (but not generally symmetrically).


\bigskip

We can alternatively  consider a simple model for $f$ with $N$ latitudinal zones, which is a generalization of the $N=2$ case used throughout this Letter. Such a model has
\[
f_{\text{zones}}(b) =
f_i \qquad \text{for } b \geq b_{i-1}  \>\>\&\>\> b < b_i \qquad (i = 1, \ldots, N)\\
\]
where $b_0, \ldots, b_{N}$ is a monotonically increasing array of zonal boundaries, with $b_0 = 0$ and $b_N = 1$, and the $f_i$ are active fractions between 0 and 1. Then
substitution in equation~\eqref{intavstar} yields
\begin{align}
\langle S \rangle 
&= S_0 + \Delta S \>
\sum_{i=1}^N f_i \left( \frac{4}{\pi} \right) 
\int_{b_{i-1}}^{b_i} dy\, (1 - y^2)^{1/2} \notag\\
&= S_0 + \Delta S \>
\sum_{i=1}^N f_i \> [ H(b_i) - H(b_{i-1}) ]
\label{Nzonemodel}
\end{align}
where we have defined the auxiliary function $H(y)$ as the indefinite integral 
\begin{equation}
H(y) \equiv \left( \frac{4}{\pi} \right) 
\int\, dy \> (1 - y^2)^{1/2}
= \left( \frac{4}{\pi} \right) \left\{ \tfrac{1}{2}y\sqrt{1-y^2} + \tfrac{1}{2}\arcsin y \right\} \quad.
\label{Hdef}
\end{equation}
This function corresponds to the fraction of the area of a unit circle that is covered by chords with absolute impact parameter $\leq y$. Hence, $\;H(0) = 0$ and $H(1) = 1$.

\bigskip

For the case of a two-zone model as discussed in this Letter ($N = 2$), equation~\eqref{Nzonemodel} reduces to:
\begin{equation} \label{eq:twozoneaverageS}
\langle S \rangle = S_0 + \Delta S\,\big[f_1 H(b_1) + f_2 (1 - H(b_1))\big]
\end{equation}
The average flux along zone $i$ is simply 
\begin{equation} \label{eq:twozoneaverageSi}
\langle S \rangle_i = S_0 + f_i \,\Delta S
\end{equation}
This implies that
\begin{equation} 
\langle S \rangle = H(b_1) \langle S \rangle_1 + (1 - H(b_1)) \langle S \rangle_2 
\end{equation}
Thus the average flux of the star is simply a linearly weighted average between the average flux of the two zones, with $H(b_1)$ being the weight.

If $H(b_1) = \tfrac{1}{2}$, then $\langle S \rangle$ is halfway between $\langle S \rangle_1$ and $\langle S \rangle_2$. Conversely, in this case $\langle S \rangle_1$ and $\langle S \rangle_2$ are both equally far from
$\langle S \rangle$. Solving equation~\eqref{Hdef} for $H(b_1) = \tfrac{1}{2}$ yields a zonal boundary at 
\begin{equation}
b_{1,\text{midpoint}} \approx 0.404 
\label{eq:bmidzone}
\end{equation}
For this specific value of the zonal boundary in a two-zone model, the TLS effect has the same absolute size for every chord, but chords below and above $b_{1,\text{midpoint}}$ have a TLS effect of opposite sign. When instead the zonal boundary is at $b_1 < b_{1,\text{midpoint}}$, then $H(b_1) < \tfrac{1}{2}$, and the TLS effect is larger in the equatorial zone 1 than in the polar zone 2. This is the case in the examples illustrated in this Letter (see Sections~\ref{sec:methods} and~\ref{sec:results}), which used $b_1= 0.259$. If instead $b_1 > b_{1,\text{midpoint}}$, then $H(b_1) > \tfrac{1}{2}$, and the TLS effect is larger in the polar zone 2 than in the equatorial zone 1. 

\vspace{5mm}

\facility{JWST (NIRSpec/PRISM)}

\software{matplotlib \citep{matplotlib}, Astropy \citep{astropy:2013, astropy:2018, astropy:2022}, expecto \citep{morris_2024_expecto}, SpectRes \citep{Carnall2017}, petitRADTRANS \citep{Molliere2019prt}
}

\bibliography{references}{}
\bibliographystyle{aasjournal}

\end{document}